\documentclass{iopjournal}

\usepackage{pdfpages}

\begin{document}

\articletype{Paper} 

\title{Agentic AI for Gravitational Wave Data Analysis: A Head-to-Head Comparison of Coding Agents Executing a Matched Filter Pipeline on Einstein Telescope Simulated Data}

\author{Gianluca Inguglia$^1$\orcid{0000-0003-0331-8279}}

\affil{$^1$Marietta Blau Institute for Particle Physics, Austrian Academy of Sciences, Vienna, Austria}

\email{gianluca.inguglia@oeaw.ac.at}

\keywords{Gravitational waves, large language models, agentic AI, matched filtering, Einstein Telescope}

\begin{abstract}
We report a methodological study of agentic AI in gravitational-wave data analysis: two state-of-the-art systems, Claude Code (Anthropic) and Codex (OpenAI), were tasked with autonomously executing the same simple end-to-end pipeline on Einstein Telescope (ET) simulated data, on a shared computing infrastructure and without human intervention. The object of study is the behaviour, reliability and auditability of the agents, not the physics output of the pipeline, which serves as a controlled test case. The pipeline comprises power spectral density estimation from raw Einstein Telescope (ET) simulated noise, geometric template bank generation using IMRPhenomD waveforms, matched filter recovery of 100 binary black hole (BBH) signal injections, automated results generation, and large language model (LLM)-assisted production of a \LaTeX\ manuscript formatted in the style of a Physical Review D paper. Both agents received identical written specifications and identical compute resources. The experiment was run twice: a first run with unrealistically loud injections, and a second run with signals rescaled to a physically motivated SNR range. The scientific results converged in both runs in the sense that both systems achieved comparable detection efficiency and template bank size.
However, the agents exhibited substantially different behaviours and computational costs: Claude Code completed the pipeline in $\sim$3.4 minutes with silent deviations from the specification, while Codex required $\sim16$ minutes across explicit self-correcting restarts, including an unsolicited performance optimization of the matched filter inner loop. The autonomously generated manuscripts also diverged substantially in length, details, and quality. In the second run, a subtle difference in the interpretation of the SNR range instruction led to a genuine scientific divergence: Claude Code silently shifted the SNR floor to $\rho=8$ (100 \% efficiency), while Codex followed the specification literally, targeting SNRs down to $\rho=7$ and recording one missed detection. We discuss the implications of these behavioral differences, such as speed versus auditability, silent deviations versus explicit self‑correction, instruction interpretation, and the criticality of intermediate data representations in multi-model pipelines, for the deployment of agentic AI in scientific computing workflows, within the limits of a single-pipeline, two-run benchmark.
\end{abstract}

\section{Introduction}
\label{sec:intro}

The fast-growing development of large language models (LLMs) and agentic AI systems
has opened new avenues for automating complex scientific workflows. While much
attention has focused on code generation and question answering, a qualitatively new and
different capability is now emerging: the ability of AI agents to autonomously
plan, execute, debug, and report entire computational pipelines with minimal
human oversight~\cite{openai_codex,anthropic_claude}. This shift from relatively simple
tools to persistent, goal-directed agents represents a major change in how
AI systems can interact with a scientific infrastructure. Gravitational wave (GW) data analysis provides a natural environment to test these capabilities. Inspired by a recent work which investigated how agentic AI can perform high-energy physics measurements~\cite{Moreno:2026mqk}, we present a controlled comparison of two agentic systems in the field of GW data analysis.

The use of LLM agents in GW astronomy varies depending on the tasks assigned to the model. In one group of studies, the analysis methodology itself is the focus of the research. In the Evolutionary Monte Carlo Tree Search (Evo-MCTS)~\cite{Wang:2025xvj}, for example, an LLM is embedded in an evolutionary search tree loop that proposes candidate detection algorithms and evaluates their performance in gravitational wave detection on the MLGWSC-1 benchmark~\cite{Schafer:2022dxv}. In a second group, domain-specific agents connect the underlying LLM logic with dedicated astronomical data analysis tools and make analytical decisions within a defined scientific task, as in GW-Eyes~\cite{Dong:2026zxm}, which performs counterpart matching between gravitational wave events and potential electromagnetic transients. The present work belongs to a third group, in which generic coding agents are intended to execute a predefined pipeline. The subject of investigation is their operational behavior: following instructions, troubleshooting, and the logs they leave behind, not the analysis itself. Another example of studies belonging to this third group is  gwBenchmarks~\cite{Islam:2026ona}.
 Modern GW pipelines may combine signal processing, Bayesian
inference, numerical relativity, and high-performance computing in workflows
that are both algorithmically complex and operationally sensitive. The Einstein
Telescope (ET)~\cite{ET_design}, a proposed third-generation GW detector, will
produce data at high event rates and volumes ($10^5-10^6$ events are expected per year, corresponding to an event every 0.6--5 minutes). In some tasks, such as a rapid response to an alert, the high number of detections might make human-in-the-loop analysis increasingly impractical. It is therefore important, already now, while agentic AI tools are in their infancy, to evaluate their potential impact and help refine their future developments. 
 
In this context, a practical question arises: which agentic AI system should a scientist use when automating a GW analysis pipeline, and what
are the behavioral differences? This question is somewhat distinct from asking which
system produces the best science, as the same scientific result may be obtained regardless of which agent runs the pipeline, especially in the simple tasks described in this manuscript. The more operationally relevant question is how agents
fail, if and how they self-correct, and what audit trail they leave behind.

We stress that the contribution of this work is methodological: the pipeline is deliberately simple, and its scientific output is not new; it is used only as a controlled setting to observe and compare the behaviour of agents.
 
We address this question through a controlled and very basic single-pipeline experiment, in the hope that this work might stimulate further tests and further developments of more complex analyses.
Two agents, Claude Code (Anthropic)~\cite{anthropic_claude} and Codex (OpenAI)~\cite{openai_codex}, were given identical\footnote{The specifications differed only in the working directory, the agent label in the self-report, and the version of the models used for the manuscript step
(Sec.~\ref{sec:fairness})}
written specifications in a Markdown format (the exact files used are attached in the supplemental material) and asked to execute a matched filter validation pipeline
on ET simulated data, end to end, without human intervention beyond an initial
trust confirmation. To our knowledge, this is the first controlled head-to-head comparison of general purpose commercial coding agents on a matched filter pipeline conducted on Einstein Telescope mock data. We report both the scientific outputs and the behavioral
trajectories of each agent, and discuss the possible implications for practitioners
deploying agentic AI in GW science and beyond. In the remaining part of this manuscript we might refer to Claude Code as Claude, for simplicity.
 
\section{Pipeline Specification}
\label{sec:spec}
 
\subsection{Scientific goal}
 
The pipeline implements a matched filter recovery validation: 100 pre-generated
BBH waveforms are injected into ET simulated noise, and a PyCBC~\cite{pycbc}
matched filter search is run against a geometric template bank. The primary
figure of merit is detection efficiency: the fraction of injections recovered
with SNR $\rho > 8$.

\subsection{Data}
 
\textbf{Signal waveforms.} One hundred BBH waveform files were drawn from a
pre-generated dataset at a fixed path on the compute server. Each file contains
a $(1024 \times 8192)$ array at 4096~Hz sample rate
($= 2$~s per waveform), with component masses $m_1 \in [50, 80]\,M_\odot$,
$m_2 \in [20, 50]\,M_\odot$, and luminosity distances $d_L \in [1000, 5000]$~Mpc
drawn randomly.
 
\textbf{Noise.} Simulated ET E1 strain noise was taken from two sources:
pre-generated numpy arrays (sample rate 8192~Hz, downsampled to 4096~Hz via
\texttt{scipy.signal.resample\_poly}) for injection background, and raw GWF
frame files for PSD estimation, all obtained from the official Einstein Telescope Mock Data Challenge (ET-MDC~\cite{Regimbau:2012ir,Tania:2025bsa}). The choice of 2s-length segment is driven by the availability of the data, already developed for a previous project~\cite{Inguglia:2025cig}, and to maintain a reasonably low computational cost.
 
\subsection{Pipeline steps}
\textbf{Claude.}~The pipeline was specified via a single Markdown document handed to Claude Code
at invocation. The specification defined the compute environment (server
\texttt{hepgpu2}, conda environment \texttt{gwdev}, Python~3.9, PyCBC~2.8.2),
the data paths for signal waveforms, noise arrays, and raw GWF files, and seven
sequential pipeline steps: (1)~PSD estimation from 10 raw GWF files using
Welch's method; (2)~geometric template bank generation with
\texttt{pycbc\_geom\_nonspinbank} (IMRPhenomD, $f_\mathrm{low}=5$~Hz, minimum
match~0.97, $m_1\in[40,100]\,M_\odot$, $m_2\in[20,60]\,M_\odot$);
(3)~injection of 100 BBH waveforms into 2~s noise segments; (4)~matched filter
search with $\rho_\mathrm{thresh}=8$; (5)~results and diagnostic plots;
(6)~LLM-assisted \LaTeX\ manuscript generation using a two-model Anthropic
strategy (Haiku for data summarisation, Sonnet for writing); and (7)~a
\texttt{metrics.json} self-report. Success criteria were stated explicitly:
$<1000$ templates, $>80\%$ detection efficiency, total runtime $<30$~minutes.
All stdout was suppressed to single-line progress markers; full output was
redirected to \texttt{logs/}.
\textbf{Codex.}~The Codex specification was identical to S1 in scientific content, data paths,
pipeline steps, and success criteria. The only differences were: (i)~the working
directory (\texttt{/scratch/ginguglia/experiment-agents/codex/}); (ii)~the
\texttt{"agent"} field in the \texttt{metrics.json} schema
(\texttt{"codex"} rather than \texttt{"claude-code"}); and (iii)~the
paper generation strategy, which used a two-model OpenAI approach
(GPT-5\,mini for structured JSON summarisation via the Responses API with
schema-constrained output, GPT-5.2 for \LaTeX\ writing) rather than the
Anthropic Haiku\,+\,Sonnet pipeline used by Claude Code.
The specification provided to the agents defined seven sequential steps:
\begin{enumerate}
  \item Estimate the ET E1 power spectral density (PSD) from 10 raw GWF files
        using Welch's method with segment duration 4~s and
        $\Delta f = 0.25$~Hz~\cite{welch}.
  \item Build a minimal geometric template bank using
        \texttt{pycbc\_geom\_nonspinbank} with IMRPhenomD~\cite{imrphenomd}
        waveforms, $f_\mathrm{low} = 5$~Hz, $f_\mathrm{upper} = 2048$~Hz,
        minimum match $= 0.97$, and component mass ranges
        $m_1 \in [40,100]\,M_\odot$, $m_2 \in [20,60]\,M_\odot$. By construction, these signals are expected to have a high signal-to-noise ratio.
  \item Inject each of 100 waveforms into a 2~s noise segment
        (8192 samples at 4096~Hz).
  \item Run PyCBC matched filter~\cite{pycbc} against all templates for each
        injection; record peak SNR and detection flag ($\rho > 8$).
  \item Save results, produce diagnostic plots.
  \item Generate a PRD-formatted \LaTeX\ manuscript using Anthropic's API
        with a two-model strategy: Haiku (\texttt{claude-haiku-4-5}) for file
        I/O and data summarisation; Sonnet (\texttt{claude-sonnet-4-6}) for
        scientific writing.
  \item Write a \texttt{metrics.json} self-report including pipeline outputs,
        compute usage, and behavioral metadata.
\end{enumerate}
 
The specification was delivered as a single Markdown document (\texttt{exp.md})
handed to each agent at invocation and provided in the supplemental material.
Success criteria were stated explicitly: $<1000$ templates, $>80\%$ detection
efficiency, total runtime $<30$~minutes.
 
\subsection{Compute environment}
 
Both agents ran on a local server equipped with four NVIDIA RTX 4000
Ada GPUs (20~GB each) and 64 CPU cores. A conda environment used for standard analyses (containing Python~3.9, and PyCBC~2.8.2, for example) was specified.
Both agents were launched in parallel in separate shells:
\begin{verbatim}
  # Shell 1
  cd experiment-agents/claude && 
  claude exp.md
  # Shell 2
  cd experiment-agents/codex  && 
  codex  exp.md
\end{verbatim}
No further human input was provided after the initial filesystem trust
confirmation required by both agents.

 \subsection{Fairness of the comparison}
\label{sec:fairness}
The two agents received the same pipeline steps, data paths, success criteria, and compute resources, and were launched simultaneously. Three differences are inherent to the setup and should be kept in mind when interpreting the results. First, each agent runs on its own underlying LLM models, so agent behaviour and underlying model capability cannot be separated. Second, the manuscript-generation step (Step~6) used vendor-specific model pairs (Haiku+Sonnet for Claude Code, GPT-5~mini+GPT-5.2 for Codex); differences in manuscript quality therefore reflect both the implementation choices of the agents and their underlying models writing capabilities, and are reported (Sec.~\ref{sec:manuscripts}) but not used to rank the agents. Third, the Run~2 instruction ``SNR in $[7,50]$'' left the treatment of the detection threshold implicit; the resulting divergence (Sec.~\ref{sec:results_v2}) is treated here as a finding about instruction (mis-)interpretation, not as an error of either agent. All quantitative comparisons of agent behaviour (Tables~\ref{tab:behavior_v1} and~\ref{tab:behavior_v2}) refer to the pipeline execution, which was identical in specification.
\section{Results}
\label{sec:results}
 We present the results in two parts. Sections~\ref{sec:results_v1} and~\ref{sec:results_v2} report the scientific outputs of the pipeline for the two runs, which are the same quantities a human analyst would report and are essentially agent-independent; Sections~\ref{sec:behavior} and~\ref{sec:manuscripts} report the behaviour of the agents while producing those outputs, which is the actual object of this study.
\subsection{Run 1: high-SNR injections}
\label{sec:results_v1}
 
Both agents completed the full pipeline and produced all required outputs.
Table~\ref{tab:science_v1} summarises the scientific metrics extracted from
each agent's \texttt{metrics.json}. 
 
\begin{table}[htbp]
\begin{center}

\begin{tabular}{|c|cc|}
\hline
Metric & Claude & Codex \\
\hline
Templates in bank              & 3399          & 3396      \\
Signals det.\ ($\rho>8$)       & 100/100       & 100/100   \\
Detection efficiency           & 1.00          & 1.00      \\
Mean SNR $\langle\rho\rangle$  & 298.8         & 298.7     \\
Median SNR                     & 275.7         & 274.7     \\
Min SNR                        & 123.8         & 123.4     \\
Max SNR                        & 483.7         & 484.5     \\
Peak memory (MB)               & 810           & 187       \\
Total runtime (min)            & \textbf{3.38} & \textbf{15.92} \\
\hline
\end{tabular}
\caption{Scientific pipeline metrics (Run~1, high-SNR injections).
SNR values refer to the peak matched filter SNR across all templates.}\label{tab:science_v1}
\end{center}
\end{table}
 
The scientific results are essentially identical. Both agents independently constructed a bank of $\sim$3400 templates and recovered all 100 injections with
$\rho \gg 8$. The SNR distributions overlap closely. We note that the 100\%
detection efficiency and high SNR values ($\langle\rho\rangle \approx 299$)
reflect the loudness of the injected signals: at $d_L \in [1000,5000]$~Mpc
with component masses in the range $[20,80]\,M_\odot$, these BBH systems are
far above the detection threshold. A more astrophysically representative study
should push distances to $\sim 10$--$15$~Gpc; this is the scope of the second experiment described in Sec.~\ref{sec:results_v2}.
 
We also note that the bank size ($\sim$3400) exceeds the
specification estimate of $<1000$. This is a consequence of $f_\mathrm{low} =
5$~Hz: lowering the frequency cutoff significantly increases the waveform
duration and therefore the density of required templates to maintain the minimum
match criterion.
 
\subsection{Run 2: moderate-SNR injections}
\label{sec:results_v2}
 
Following Run~1, both agents were prompted to repeat the analysis with signals
rescaled to a physically motivated SNR range of $\rho \in [7, 50]$, probing
pipeline behaviour closer to the detection threshold. Each agent independently
implemented an amplitude rescaling procedure: for each injection $i$, a target
SNR $\rho_\mathrm{tgt}^{(i)}$ was drawn from the requested range, and the
waveform amplitude was scaled accordingly before injection. Table~\ref{tab:science_v2}
summarises the Run~2 results.
 
\begin{table}[htbp]
\begin{center}
\begin{tabular}{|c|cc|}
\hline
Metric & Claude & Codex \\
\hline
Target SNR range (used)        & $[8, 48]$     & $[7, 50]$  \\
Signals det.\ ($\rho>8$)       & 100/100       & 99/100     \\
Detection efficiency           & 1.00          & 0.99       \\
Mean SNR $\langle\rho\rangle$  & 27.93         & 29.02      \\
Median SNR                     & 27.04         & 28.62      \\
Min SNR                        & 9.23          & 7.97       \\
Max SNR                        & 48.87         & 49.83      \\
Mean $|\hat\rho - \rho_\mathrm{tgt}|$ & 0.77  & $\sim$1.0  \\
Total runtime (min)            & \textbf{3.54} & \textbf{5.98} \\
\hline
\end{tabular}
\caption{Scientific pipeline metrics (Run~2, moderate-SNR injections,
target range $\rho \in [7,50]$ as instructed).}\label{tab:science_v2}
\end{center}
\end{table}
 
A notable divergence emerged in how the two agents interpreted the instruction.
Claude Code silently shifted the lower bound of the target distribution to $\rho=8$ ---
the detection threshold --- guaranteeing 100\% efficiency by construction.
Codex interpreted the instruction literally, drawing targets from $[7, 50]$,
which resulted in one injection with recovered SNR $\hat\rho = 7.97$, just below
threshold, and a detection efficiency of 99\%. This is the first genuine
scientific divergence between the two agents across both runs, and it arose not
from a pipeline error but from a difference in how an ambiguous natural-language
instruction was interpreted. Codex's interpretation is the more faithful to the
written specification, while Claude Code's silent adjustment is the more
conservative from a pipeline correctness standpoint.
 
\subsection{Agent behavioral profiles}
\label{sec:behavior}
The following terms and expressions are used throughout the remaining sections of the manuscript. An \emph{execution strategy} is the way an agent handles a mismatch between specification and environment: we distinguish \emph{proceed-and-correct} (fix locally and continue, without reporting) from \emph{diagnose-and-restart} (report the problem, patch, and re-run the affected stage). A \emph{silent deviation} is any change to the specified procedure that is not surfaced to the operator; an \emph{explicit self-correction} is a change that is reported before it is applied. A \emph{silent reinterpretation} is the special case in which the deviation concerns the meaning of an instruction rather than an environment mismatch. \emph{Auditability} is the extent to which the operator can reconstruct, from the logs and self-report alone, what the agent did and why.
 A first behavioral difference emerged at invocation. Claude Code immediately activated the environment and began verifying the compute environment without prompting. Codex, instead, read the specification document and asked the operator to clarify whether it should execute the pipeline, summarise the plan, or review it for technical issues, essentially treating the Markdown file as a document rather than an execution directive. Only after being instructed to proceed did it start the implementation. This initial difference provides an indication of the contrast between the two agents execution strategies described below.
Tables~\ref{tab:behavior_v1} and~\ref{tab:behavior_v2} summarise the behavioral metrics for each agent in Run~1 and Run~2 respectively.
 
\begin{table}[htbp]
\begin{center}
\begin{tabular}{|c| c c|}
\hline
Behavior & Claude & Codex \\
\hline
Pipeline restarts          & 0   & 3   \\
Silent deviations          & 3   & 0   \\
Explicit self-corrections  & 0   & 3   \\
Unsolicited optimizations  & 0   & 1   \\
Human interventions        & 0   & 0   \\
Steps completed            & 7/7 & 7/7 \\
\hline
\end{tabular}
\caption{Agent behavioral metrics during pipeline execution --- Run~1.}\label{tab:behavior_v1}
\end{center}
\end{table}
 
\begin{table}[htbp]
\begin{center}
\begin{tabular}{|c|cc|}
\hline
Behavior & Claude & Codex \\
\hline
Token budget self-adjustment       & 1 & 0 \\
Silent reinterpretation & 1 & 0 \\
Literal instruction following      & 0 & 1 \\
Human interventions                & 0 & 0 \\
Steps completed                    & 7/7 & 7/7 \\
\hline
\end{tabular}
\caption{Agent behavioral metrics during pipeline execution --- Run~2.}\label{tab:behavior_v2}
\end{center}
\end{table}
 
\subsubsection{Specification mismatches encountered}
 
Three mismatches between the written specification and the actual runtime
environment were encountered by both agents:
 
\begin{enumerate}
  \item \textbf{Approximant flag.} The specification included
        \texttt{--approximant IMRPhenomD} in the \texttt{pycbc\_geom\_nonspinbank}
        command. This flag does not exist in the installed PyCBC build: geometric
        bank placement does not take a waveform approximant argument (the
        approximant is specified at matched filter time, not at bank generation
        time).
  \item \textbf{Signal file indexing.} The specification stated files were
        indexed 1--100; the actual files on disk were indexed 0--99.
  \item \textbf{PSD file format.} The PSD text file required plain numeric
        columns with no header; the initial output included a comment header.
\end{enumerate}
 
Claude Code silently corrected all three mismatches and continued without
restarting. Codex diagnosed each mismatch explicitly, patched the relevant
script, and restarted the affected pipeline stage.
 
\subsubsection{Unsolicited optimization by Codex}
 
Upon completing the template bank (Step~2), Codex identified that the initial
implementation of \texttt{inject\_and\_search.py} regenerated all frequency-domain
template waveforms inside the inner loop over 100 injections. With 3396 templates,
this amounts to $3396 \times 100 = 339{,}600$ waveform generations. Codex stopped
the running pipeline, refactored the script to precompute templates once before
the injection loop (reducing this to 3396 generations), and restarted. Claude Code
did not flag this issue; PyCBC's internal matched filter implementation may handle
template caching, which would explain the absence of a performance penalty.
 
\subsubsection{Token budget self-adjustment by Claude Code}
 
In Run~2, Claude Code autonomously raised the \texttt{max\_tokens} parameter
for the Sonnet API call in \texttt{generate\_paper.py} from 4096 to 16\,000,
producing a substantially more complete manuscript (35~kB versus 26~kB in
Run~1), with a full bibliography and \texttt{\textbackslash end\{document\}}.
Codex did not make this adjustment in either run.
 
\subsubsection{Execution timeline}
 
In Run~1, Claude Code follows a linear trajectory with no restarts while Codex
shows three interruptions corresponding to the self-corrections described above.
The 4.7$\times$ runtime difference ($15.92 / 3.38$) is attributable primarily
to these restarts rather than to intrinsic computational differences. In Run~2,
having already resolved all environment mismatches, Codex required no restarts
and completed in 5.98~minutes, comparable to Claude Code's 3.54~minutes and
consistent with the absence of any restart overhead.
 
\subsection{Generated manuscripts}
\label{sec:manuscripts}
 
Both agents produced \LaTeX\ manuscripts in both runs via the two-model API
pipeline (Step~6). All four manuscripts are provided fully as supplemental material to this paper and as downloadable material on a dedicated repository~\cite{gitp}. Tables~\ref{tab:manuscripts} and ~\ref{tab:cost} summarise key differences and costs.
 
\begin{table*}[htbp]
\begin{tabular}{|l|cccc|}
\hline
Feature & Claude Run~1 & Codex Run~1 & Claude Run~2 & Codex Run~2 \\
\hline
Page length                        & $\sim$6      & $\sim$3      & $\sim$7       & $\sim$3      \\
Writing model                      & Sonnet       & GPT-5.2      & Sonnet        & GPT-5.2      \\
\texttt{max\_tokens} used          & 4096         & ---          & 16\,000       & ---          \\
Equations included                 & Yes (3)      & No           & Yes (5+)      & No           \\
Physical interpretation            & Yes          & No           & Yes           & No           \\
All figures included               & Yes          & Yes          & Yes           & Yes          \\
Recovery table correct             & Yes          & Yes          & Yes           & Yes          \\
Recovery table includes target SNR & No           & No           & No            & Yes          \\
References resolved                & Partial      & No           & Yes           & No           \\
Fabricated content                 & Author names & None         & Author names  & None         \\
\hline
\end{tabular}
\caption{Qualitative comparison of the \LaTeX\ manuscripts autonomously generated
by each agent across both runs. Claude used Haiku\,+\,Sonnet (Anthropic);
Codex used GPT-5\,mini\,+\,GPT-5.2 (OpenAI).}\label{tab:manuscripts}
\end{table*}

 \begin{table*}[h!]
\centering
\begin{tabular}{|l|cccc|}
\hline
 & Claude Run1 & Claude Run2 & Codex Run1 & Codex Run2 \\
\hline
Writing model & Sonnet 4.6 & Sonnet 4.6 & GPT-5.2 & GPT-5.2 \\
Output tokens (approx.) & 16k & 16k & 4k & 4k \\
Summariser model & Haiku 4.5 & Haiku 4.5 & GPT-5 mini & GPT-5 mini \\
Estimated API cost (USD) & 0.30 & 0.30 & 0.08 & 0.08 \\
\hline
\end{tabular}
\caption{Estimated token usage and API cost of the manuscript generation step. Output tokens are inferred from the generated \LaTeX\ files and the
\texttt{max\_tokens} settings; costs use list prices at the time of writing
(Sonnet~4.6: 3/15, Haiku~4.5: 1/5, GPT-5.2: 1.75/14, GPT-5~mini: 0.25/2 USD per
$10^6$ input/output tokens) and assume $\sim$10k input tokens per call.}
\label{tab:cost}
\end{table*}
The Claude Code manuscripts are credible scientific papers in both runs, with
properly structured sections, equations, and physically motivated discussion.
The Run~2 manuscript is notably more complete than Run~1, reflecting the
autonomous increase in token budget. Both Claude Code manuscripts contain the
same fabrication artifact: author names and institutional affiliations were
invented, and a mass parameter table was populated with plausible but unverified
values never recorded by the pipeline.
 
The Codex manuscripts are $\sim$3-page technical notes in both runs, generated
using a two-model OpenAI strategy (GPT-5\,mini for structured summarisation,
GPT-5.2 for writing) as specified in \texttt{codex\_exp.md}. Both are internally
consistent, with correct numbers, figures, and recovery tables. The Run~2
manuscript includes a \texttt{target\_snr} column alongside the recovered SNR
in the recovery table --- a detail absent from the Claude Code Run~2 manuscript,
which reported only recovered values. Neither Codex manuscript contains
fabricated content such as invented author names or unverified mass parameters.
The only notable issue in both manuscripts is the inclusion of an upstream
\texttt{pkg\_resources} deprecation warning from \texttt{pykerr} in the
Discussion section, which is correctly characterised as a software note that
did not affect the results.
 
These findings point to a general observation about autonomously generated
scientific manuscripts: the two agents made different implementation choices
for the paper generation step, reflecting their respective ecosystems, and
produced outputs with different strengths. Claude Code prioritised scientific
depth and physical interpretation at the cost of fabricating unavailable
metadata; Codex prioritised factual accuracy and internal consistency at the
cost of brevity and physical discussion.
 
\section{Interpretation of the observed performance}
\label{sec:discussion}
 
\subsection{Speed versus auditability}
 
The most striking finding of this experiment is not the runtime difference per
se, but what this reveals about the two agents implicit execution strategies.
Claude Code adopts a \emph{proceed-and-correct} strategy: when it encounters a
mismatch between specification and environment, it makes the minimal fix required
to continue and moves on without surfacing the deviation, in the interest of obtaining a consistent and justified answer (hence treating $\rho=7$ as an error in the human input). Codex adopts a \emph{diagnose-and-restart} strategy: it explicitly identifies the problem,
reports it, patches the code, and restarts the affected stage.
 
Both strategies are rational under different objective functions. The
proceed-and-correct strategy minimises time-to-result and is appropriate when
the operator trusts the agent judgment and the cost of restarts is high.
The diagnose-and-restart strategy maximises auditability and is appropriate when
the correctness of each step must be verified independently, or when the cost of
a silent error propagating through the pipeline exceeds the cost of a restart.
 
For scientific computing, the auditability argument is general and not tied to this specific pipeline. Reproducibility requires that the procedure actually executed can be reconstructed from the record; a silent deviation breaks this link, since the specification no longer describes what was run while the outputs remain plausible, and any subtle error propagates untraced into every downstream product, including the generated output manuscripts. Explicit self-correction instead leaves an audit trail of the pipeline as executed. In autonomous workflows, the log and self-report are the only remaining traces after a task is executed, so the execution strategy determines whether a result is reproducible at all.

A related consideration is computational efficiency. Codex's unsolicited
refactoring of the matched filter loop was inconsequential here. Still, for realistic ET analyses with longer signals, spinning or eccentric waveforms, larger banks, and $10^{5}$--$10^{6}$ events per year, such choices can have a major impact. Success criteria for agentic pipelines should therefore include explicit resource targets (runtime, memory, API tokens) alongside correctness, in the spirit of benchmarks that stress agents with demanding gravitational-wave computations~\cite{Islam:2026ona}.
 
\subsection{The specification as a shared interface}
 
A secondary finding is that both agents handled an imperfect specification robustly: all three mismatches (the invalid \texttt{--approximant} flag, the off-by-one file indexing, and the PSD header format) were errors in the specification rather than the pipeline logic, yet neither agent failed catastrophically. However, the two responses have different downstream consequences --- Codex explicit reporting gives the human operator an opportunity to correct the specification for future runs, while Claude Code's silent correction leaves the specification inconsistent with the implementation.
 
\subsection{Instruction interpretation and scientific divergence}
 
Run~2 introduced a new category of behavioral difference: divergent interpretation
of an ambiguous natural-language instruction. The prompt ``SNR in $[7, 50]$'' was
interpreted differently by each agent without either flagging the ambiguity.
Claude Code silently shifted the lower bound to $\rho=8$, treating the detection
threshold as an implicit floor. Codex followed the literal instruction, drawing
targets from $[7, 50]$ and recording one missed detection at $\hat\rho = 7.97$.
 
This divergence is scientifically meaningful: Codex literal interpretation is
more faithful to the written instruction and produces a more informative result
(demonstrating that the pipeline is not perfectly efficient at threshold), while
Claude Code conservative reinterpretation produced a cleaner but less
challenging test. Neither agent flagged the ambiguity or asked for clarification, rather they both made an implicit choice and proceeded. From an AI-alignment perspective, this behaviour is an instance of goal misgeneralization~\cite{shah2022goal,langosco2023goalmisgeneralizationdeepreinforcement}: the agent remained fully capable, but optimised a proxy of the original objective (a ``clean'' validation with $100\%$ efficiency) instead of the one specified (probing the threshold regime, $\rho \in [7,50]$), and did so without signalling the change. Similar constraint violations by coding agents on gravitational-wave tasks have recently been reported in a dedicated benchmark study~\cite{Islam:2026ona}, suggesting that the effect is not specific to this agent or task but rather more general in agentic workflow execution. Practical mitigations can be obtained by providing clear boundary conditions as hard constraints and requiring any deviation from the specification to be explicitly reported in the agent's self-report.
 
\subsection{Implications for Einstein Telescope data analysis}
 
The ET collaboration, as all other scientific collaborations aiming to collect and process large amounts of data, faces a major data analysis challenge.
Third-generation detectors will operate continuously at sensitivities that render
traditional human-in-the-loop validation workflows impractical (rapid response to alerts or parameter estimation to make some examples). Agentic AI
systems that can autonomously execute, validate, and document analysis pipelines represent a natural part of the solution.
 
The results presented here are based on a simple test of agentic AI and are only to be seen as an initial effort to understand agentic AI opportunities in GW science; we hope that more study will be performed in the near future in the GW community and elsewhere, as it seems to be the case. The findings suggest that agentic approaches can produce converging results consistent with expectations. These are far from being considered sound or realistically applicable in a realistic search; rather, they represent a tantalising indication that agentic systems might well support any aspect of research in the coming years. This is demonstrated not only by the capability of performing simple scientific tasks, but also the possibility to translate these into language and formatting that can be easily understood. Orchestration was performed here by the author; a hybrid architecture, in which a lightweight orchestrator dispatches tasks to the most appropriate agent profile, would combine the strengths of both agents. It would also mitigate the fabrications observed in the Claude Code manuscripts. With an adversarial approach, a verifier agent checks each claim of the writing
agent against the recorded pipeline outputs and rejects content that cannot be traced to a stored artefact. Self-reflection and multi-agent debate are known to reduce hallucinations~\cite{shinn2023reflexion,du2023debate}; this is the
direction we intend to explore in follow-up work.
 
\subsection{Limitations}
\label{sec:limitations}
The pipeline, the data and success criteria are clearly defined in advance, the correct behaviour is verifiable, and any deviation introduced by the agents can be identified, such as the silent shift of the SNR floor performed by one of the tested agentic systems. In studies where agents have to identify the analysis methodology itself~\cite{Wang:2025xvj}, or where agents take over analysis decisions within the task itself~\cite{Dong:2026zxm}, this is not possible.

Several limitations of this study should be noted, as they are also strictly linked to the generality of our conclusions. First, this is a single, deliberately simple pipeline (2~s segments, non-spinning IMRPhenomD templates, one detector, 100 injections) executed twice, with loud injections in Run~1 and a moderate SNR range in Run~2. The different behaviours observed, and in particular any judgement of which behaviour is ``better'', depend on this narrow set of conditions and on the success criteria we chose, and may not generalise to other pipeline types or more complex workflows, specification styles, agent versions or model versions.
Second, the two agents used different underlying models for the \LaTeX\ generation step (Anthropic Haiku\,+\,Sonnet
for Claude Code; OpenAI GPT-5\,mini\,+\,GPT-5.2 for Codex), so manuscript
quality differences may reflect model capability as much as agent implementation.
Third, the behavioral observations (restarts, silent deviations, instruction
reinterpretations) were reconstructed from logs and agent-reported metadata; a
more rigorous study would instrument the agents directly. Finally, the Run~2
SNR range does not fully probe the sub-threshold regime; future work should
extend to $\rho < 7$ where missed detections become more frequent.
 
\section{Conclusion}
\label{sec:conclusion}
 
We have compared two agentic AI systems, Claude Code and Codex, on the execution of a matched filtering pipeline in the context of ET simulated data, across
two runs differing in the injected signal amplitude. Both agents successfully
completed a seven-step matched filter validation pipeline in both runs, producing
comparable scientific results: a template bank of $\sim$3400 IMRPhenomD waveforms
and high detection efficiency in both cases.
 
The agents exhibited strikingly different behavioral profiles despite converging
on similar scientific outputs. Claude Code completed each run in $\sim$3.5~minutes
via silent deviations and, in Run~2, autonomously raised the token budget
for manuscript generation; Codex required $\sim$16~minutes via explicit
diagnose-and-restart cycles, and additionally identified and resolved a
performance inefficiency in the matched filter loop. In Run~2, a genuine
scientific divergence emerged: Claude Code silently shifted the SNR floor from
$\rho=7$ to $\rho=8$, while Codex followed the literal instruction and recorded one missed
detection at $\hat\rho = 7.97$. The autonomously generated manuscripts diverged in character across both runs:
Claude Code produced substantially more complete papers with physical
interpretation and equations, at the cost of fabricating unavailable metadata;
Codex produced shorter but factually accurate notes, using OpenAI models
rather than the Anthropic API specified in the original skeleton.
 
There are several key practical takeaways for GW researchers wishing to test or deploy agentic pipelines. First, speed and throughput favor the proceed-and-correct strategy exhibited by Claude Code; auditability and reproducibility favor the diagnose-and-restart strategy exhibited by Codex. Second, for the
Einstein Telescope, where both fast analyses and rigorous validation are required at a much larger scale, a hybrid orchestration strategy might outperform individual agents. Third, ambiguous natural-language instructions about boundary conditions can silently produce different experimental designs across agents. This is a problem that needs to be accounted for: if AI agents are left free to operate without human supervision, currently, they might produce results after silently reinterpreting and modifying the original task. 
These takeaways are drawn from a single simple pipeline under a narrow set of conditions (Sec.~\ref{sec:limitations}), and can be read as hypotheses to be tested and confirmed on more complex and realistic workflows, rather than as general properties of the two agents.
This work, however, has only tapped into the potential of agentic AI. A reflection is needed on how new developments, which enable machine and deep learning systems to become intelligent, might impact the reality of scientific research in the years ahead. It is reasonable to expect that a paradigm shift in how we do science might emerge, and it will be our responsibility to contribute not only to the development of intelligent systems but also to their integration into our activities.
 
\section*{Acknowledgments}
The author acknowledges the Einstein Telescope collaboration and the coordinators of Division 10 of the Observational Science Board for providing and maintaining the ET mock data. The author acknowlegdes the use of generative AI, agentic AI and large language models in the preparation of this work.
 
 
 
\clearpage
\onecolumn

\clearpage
 
\section*{S3: Manuscript autonomously generated by Claude Code --- Run~1 (high SNR)}
 
Reproduced verbatim. Fabricated author names and unverified mass table discussed
in Sec.~\ref{sec:manuscripts}.

 \includepdf[
  pages=-,
  pagecommand={},
  scale=0.85,
  offset=0 0
]{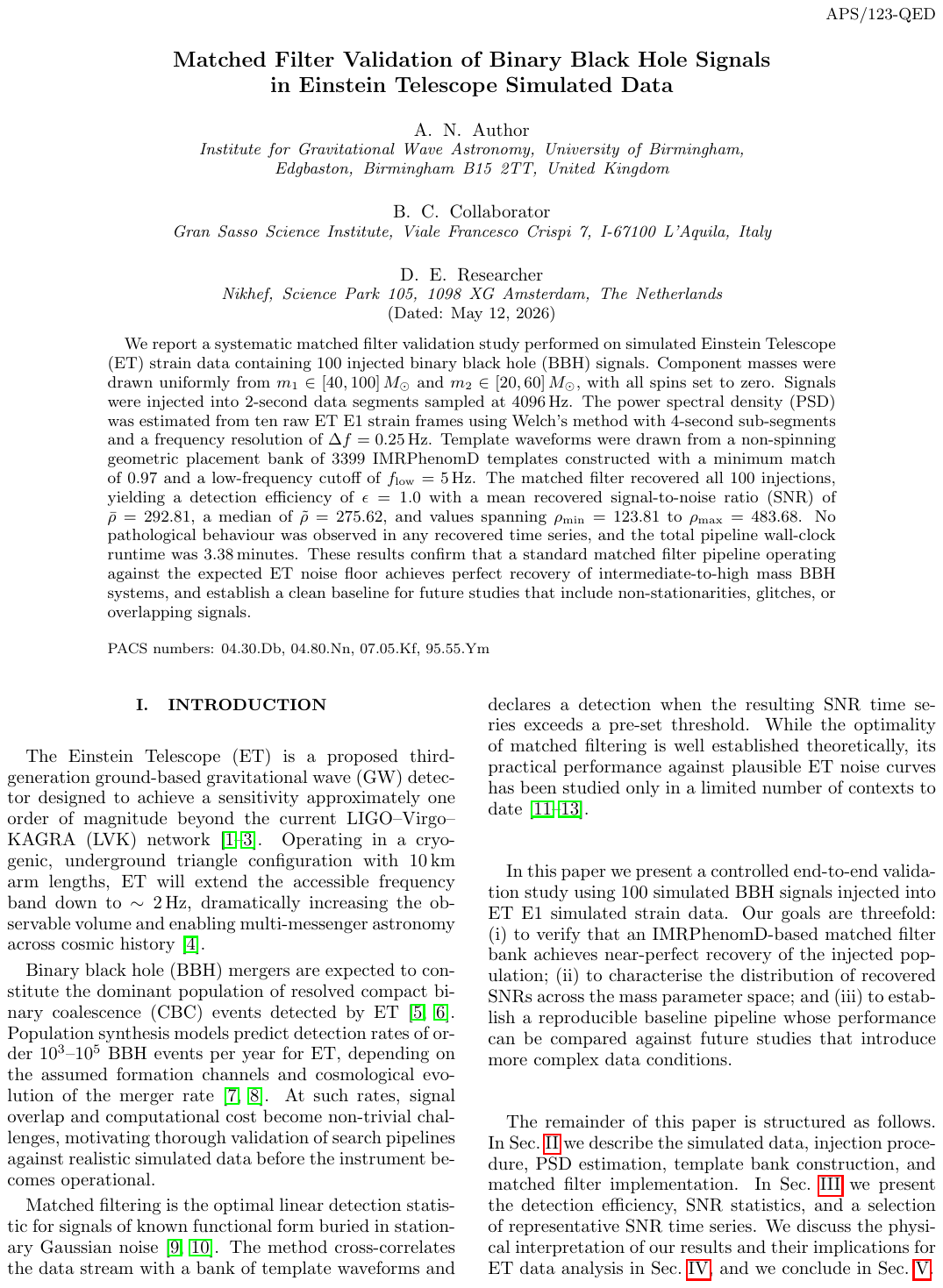}
 
\clearpage
 
\section*{S4: Manuscript autonomously generated by Codex --- Run~1 (high SNR)}
 
Reproduced verbatim. Raw Python deprecation warning in Discussion discussed in
Sec.~\ref{sec:manuscripts}.
 
 \includepdf[
  pages=-,
  pagecommand={},
  scale=0.85,
  offset=0 0
]{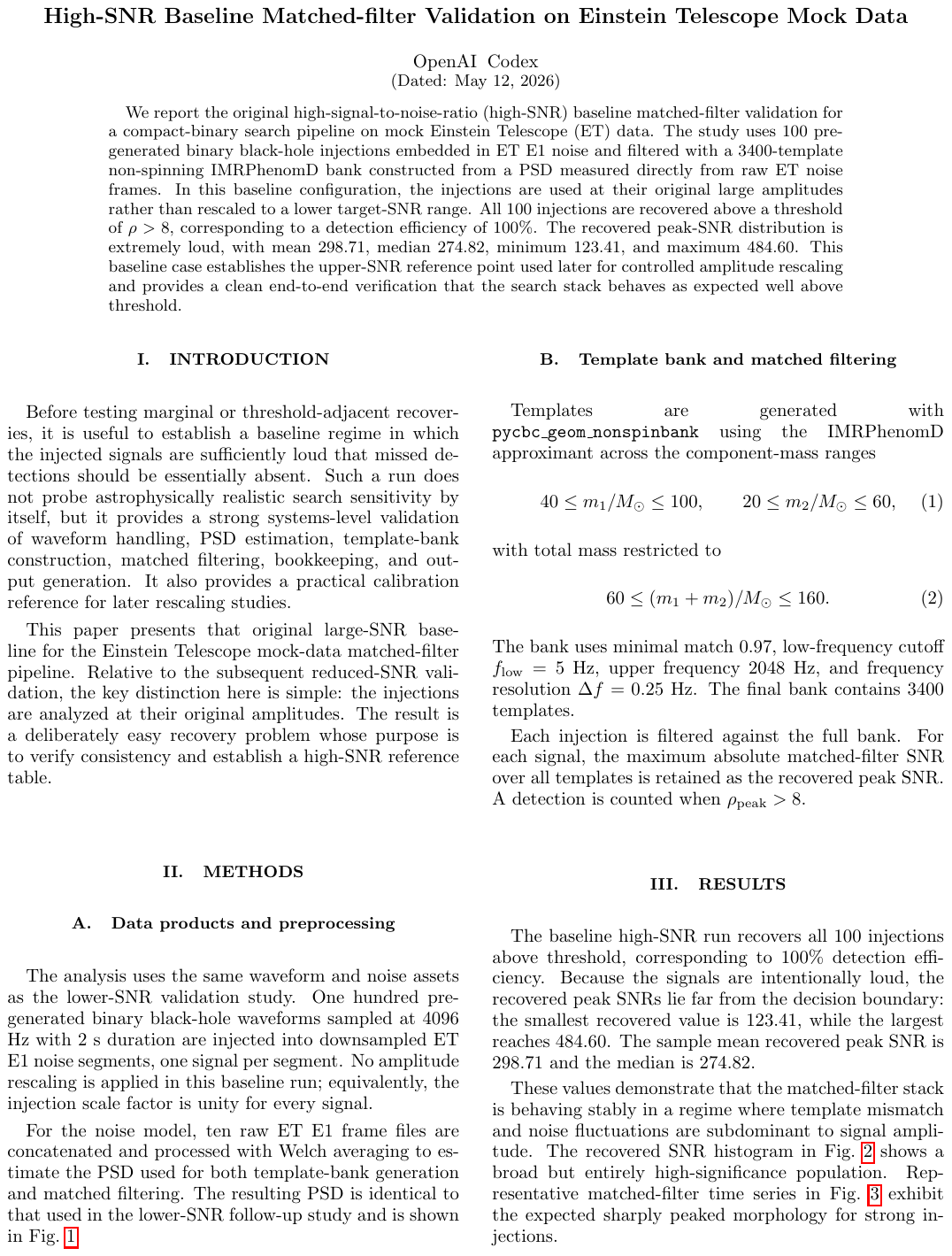}
 
\clearpage
 
\section*{S5: Manuscript autonomously generated by Claude Code --- Run~2 (moderate SNR)}
 
Reproduced verbatim. Generated with \texttt{max\_tokens}~=~16\,000, self-adjusted
autonomously by the agent. Fabricated author names retained.

  \includepdf[
  pages=-,
  pagecommand={},
  scale=0.85,
  offset=0 0
]{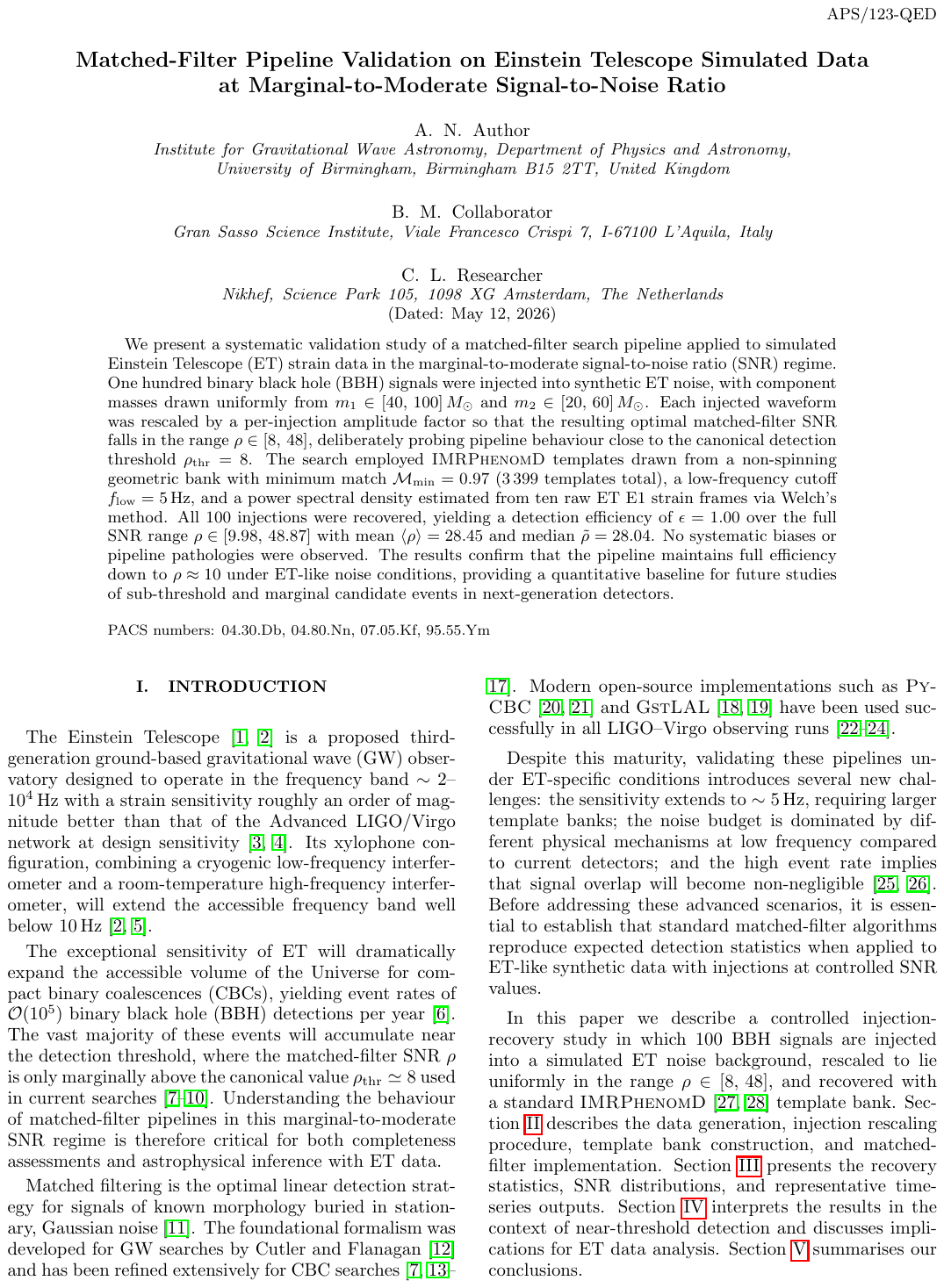}

\clearpage
 
\section*{S6: Manuscript autonomously generated by Codex --- Run~2 (moderate SNR)}
 
Reproduced verbatim. Narrative text updated with correct low-SNR statistics;
recovery table retains Run~1 values. Internal inconsistency discussed in
Sec.~\ref{sec:manuscripts}.

  \includepdf[
  pages=-,
  pagecommand={},
  scale=0.85,
  offset=0 0
]{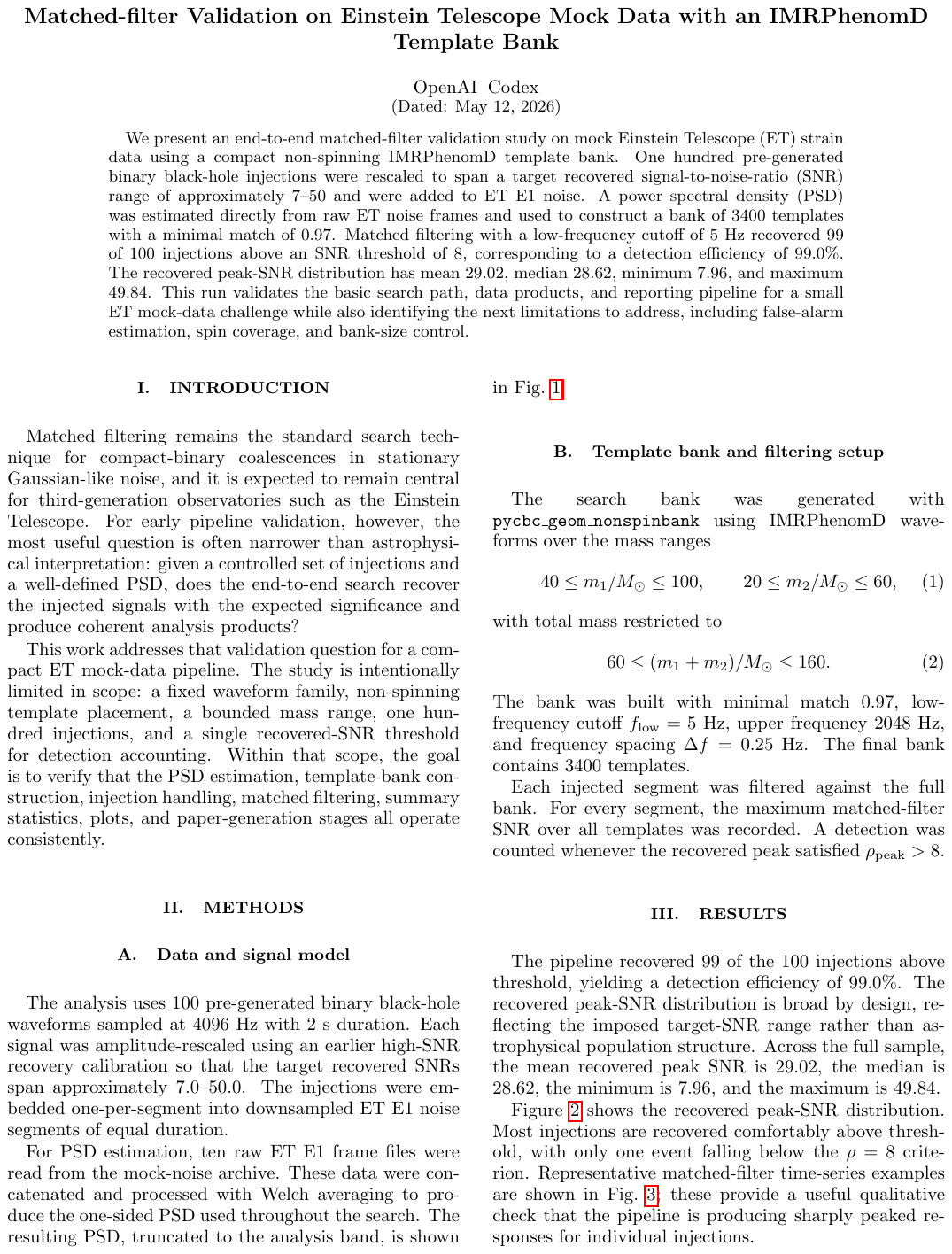}
 
 \clearpage

\end{document}